\documentclass[10pt,preprint]{aastex}

\slugcomment{To appear in Proc. SPIE}

\shorttitle{Telescope Aperture Cost Scaling Relationship}

\shortauthors{van Belle, Meinel and Meinel}

\begin{document}


\title{The Scaling Relationship Between Telescope Cost and Aperture Size
for Very Large Telescopes}


\author{Gerard T. van Belle\altaffilmark{1}}
\affil{Michelson Science Center, California Institute of
Technology, Pasadena, CA 91125\\
gerard@ipac.caltech.edu}

\and

\author{Aden Baker Meinel\altaffilmark{2,3} \& Marjorie Pettit Meinel\altaffilmark{ 3}}
\affil{Jet Propulsion Laboratory, California Institute of
Technology, Pasadena, CA 91109\\
ameinel@earthlink.net}


\altaffiltext{1}{For reprints, please contact:
gerard@ipac.caltech.edu.}
\altaffiltext{2}{OSA Fellow}
\altaffiltext{3}{SPIE Fellow}


\begin{abstract}

Cost data for ground-based telescopes of the last century are analyzed for
trends in the relationship between aperture size and cost.
We find that for apertures built prior to 1980, costs scaled as
aperture size to the 2.8 power, which is consistent with the previous finding of
Meinel (1978). After 1980, `traditional' monolithic mirror
telescope costs have scaled as aperture to the 2.5 power.
The large multiple mirror telescopes built or in construction during
this time period (Keck, LBT, GTC) appear to deviate from this relationship with significant cost
savings as a result, although it is unclear what power law such structures follow.
We discuss the implications of the current cost-aperture size data
on the proposed large telescope projects of the next ten to twenty years.
Structures that naturally tend towards the 2.0 power
in the cost-aperture relationship will be the favorable choice
for future extremely large apertures; our expectation is that
space-based structures will ultimately gain economic
advantage over ground-based ones.

\end{abstract}

\section{Introduction}

The most basic parameter that can be used to describe a telescope is its primary aperture
size.  In many cases, that parameter is such an inseparable part of a telescope's identity,
it is in fact part of the name, or at least cited in the same breath as its proper name - the 5-m Hale,
the 3.5-m WIYN, etc.
As first explored by \citet{mei78,mei79a,mei79b}, this parameter can be linked to another fundamental
parameter - that of cost.
Many additional parameterizations can be utilized to specific the capabilities and performance of a
given telescope, such as moving mass, instrument suite, and site, but aperture size
is matched only by choice of operational wavelength as a fundamental cost driver.  As described
in \citet{mei78}, a proportionality of cost to aperture size that scales as the 2.8 power
(cost $ \propto D^{2.8}$) was
found to be true to first order.
In this manuscript, we will explore the impact that
an entire generation of telescopes since then has had upon the aperture-cost power law.

\section{Ground-Based Telescopes}

Data used in this analysis can be found in Table 1, and their online cost references
\& other backing information; we have made every effort to obtain the
publicly published cost data
that most accurately reflect the telescope construction cost.
For each telescope, the cost data point was intended to be inclusive of telescope
mirror, structure, enclosure, and other essential
site work based on these references, and exclusive of instrumentation
and operations cost.
Cost data were normalized to year 2000 US dollars using the standard federal
tables for inflation adjustment for the past century.  There are unfortunately as many
acronyms as there are telescopes, and rather than expand them all here,
the reader is encouraged to reference those links he or she has an interest in.

We should note that the costs cited herein are potentially a bit `soft', in that in many cases,
a telescope's initial construction is followed by a period (sometimes years) in which the
operation of the aperture is optimized.  In many cases this optimization is improving
the performance of the telescope beyond its initial specifications, but in a few
cases this commissioning phase is needed just to meet the original design goals.  For
that latter case, the operation costs of that extended commissioning phase should
be included in the true aperture cost, but we are unable to precisely do such accounting
here.

\subsection{Pre-1980}

Large telescopes built prior to 1980 had certain basic characteristics typically in common.  These
characteristics include:

\begin{itemize}
\item Equatorial mounts - Even the massive 5-m Hale has a equatorial mount, with an axis parallel to the
Earth's rotational axis.  The period 1970-1980 saw the first breaks with tradition on this point, with the
6-m Soviet (now Russian) SAO telescope.
\item Slow optical systems - F/ratios were typically greater than 3, and never less than 2.5.
\item Thick mirrors - Some lightweighting was incorporated into these mirrors, but thermal inertia
and the resultant mirror seeing remains a substantial problem for these apertures.
\end{itemize}
As a result of the first two points above, such
designs had substantial impact upon tube length, and as a result, enclosure size and
attendant expense.

\subsection{Post-1980}

Large telescopes built after 1980 had the following basic characteristics in common:

\begin{itemize}
\item Alt-az mounts - Advances in computer control and optomechanical devices now allow for the more compact
mounting allowed by the alt-ax mounts
\item Fast optical systems - Of the major ($>$2.5m) apertures built since 1980, not a single one had a
f/ratio greater than 2.5, and none since 1989 have been greater than 1.8.
\item Thin mirrors - Often these primaries are coupled with active control systems to dynamically
compensate for changes in the angle between the pointing vector and the gravity vector.
\end{itemize}

A special class of telescopes in the post-1980 era are the {\it giant segmented mirror} (GSM)
{\it telescopes}.  Beginning with the Keck I telescope, optical systems in excess of 8.4-m have begun to be available
to the astronomical community.  Currently operational GSM telescopes are Keck I, Keck II, and the
Hobby-Eberly telescope, with the GTC, SALT, and LBT apertures all under construction.  These
telescopes all have effective areas in excess of 9-m.

A second special class appearing in the post-1980 era were large telescopes that
made special efforts in trading operation capability for increased aperture size.  Both the
Hobby-Eberly and SALT telescopes have eliminated structural elevation pointing for
simplified design and reduced cost, and the liquid mercury telescope of Univ. British Columbia
is restricted to zenith pointing for even greater cost savings, much like the Arecibo 305
meter radio telescope.

\begin{deluxetable}{llccccccl}\label{tab_telescope_data}
\rotate
\tablecolumns{9} \tabletypesize{\scriptsize} \tablewidth{0pc}
\tablecaption{Telescope data used in this analysis.\label{table1}}
\tablehead{
 \colhead{Telescope}
&\colhead{Institute}
&\colhead{Size}
&\colhead{Cost}
&\colhead{Year}
&\colhead{Adj. Cost}
&\colhead{f/ratio}
&\colhead{Mass}
&\colhead{Reference}\\
 \colhead{}
&\colhead{}
&\colhead{(m)}
&\colhead{(\$M)}
&\colhead{}
&\colhead{(2000, \$M)}
&\colhead{}
&\colhead{(tons)}
&\colhead{}
}
\startdata

Yerkes & Univ. Chicago & 1.0 & 0.5 & 1897 & 10.0 &   &  & http://www2.uchicago.edu/alumni/alumni.mag/9702/9702Yerkes.html \\
Hooker & Mt. Wilson & 2.5 & 0.6 & 1917 & 9.2 & 5.0 &  & http://www.sierramadre.lib.ca.us/smarchives/Exhibits\_More.htm \\
Hale & Caltech & 5.1 & 6.3 & 1928 & 60.0 & 3.3 & 482 & http://www.astro.caltech.edu/observatories/palomar/history/\\
&&&&&&&&http://www.gryp.fsnet.co.uk/ast11.htm \\
Mayall & NOAO & 4.0 & 10.0 & 1970 & 41.5 & 2.8 & n/a & http://www.aura-nio.noao.edu/book/ch5/5\_7.html \\
AAT & AAO & 3.9 & 22.8 & 1973 & 78.8 &   &  & http://www.ast.cam.ac.uk/AAO/about/aat.html; A\$ 15932250 \\
Blanco & CTIO & 4.0 & 10.0 & 1976 & 27.1 & 2.8 & 310 & http://arjournals.annualreviews.org/doi/full/10.1146/annurev.astro.39.1.1 \\
ESO 3.6m & ESO & 3.6 & 41.7 & 1977 & 104.3 & 3.0 & 240 & http://www.eso.org/gen-fac/pubs/astclim/papers/lz-thesis/node23.html\\
&&&&&&&&http://astrophysics.weber.edu/Correct/Astr0179.pdf \\
IRTF & NASA & 3.0 & 10.0 & 1979 & 21.7 & 2.5 &  & http://www.hawaii-county.com/databook\_97/section13.htm \\
CFHT & CFHT Consortium & 3.6 & 30.0 & 1979 & 65.1 & 3.8 &  & http://www.hawaii-county.com/databook\_97/section13.htm \\
WHT & Obs. Roque de los Muchachos & 4.2 & 21.5 & 1979 & 46.6 & 2.5 & 210 & http://www.ing.iac.es/$\sim$crb/wht/hist.html; 10 million pounds \\
Faulkes & UH / UK & 2.0 & 6.0 & 2001 & 5.9 &   &  & http://www.bizjournals.com/pacific/stories/2001/03/26/focus4.html \\
NOT & NOTSA & 2.5 & 4.9 & 1983 & 7.9 & 2.0 & 43 & http://www.not.iac.es/ \\
NTT & ESo & 3.5 & 13.0 & 1988 & 17.6 & 2.2 & 110 & http://astrophysics.weber.edu/Correct/Astr0179.pdf \\
ARC & Apache Point Obs. & 3.5 & 11.0 & 1988 & 14.9 & 1.8 &  & http://www.eurekalert.org/pub\_releases/2001-09/uoca-uoc090601.php\\
&&&&&&&&http://archives.thedaily.washington.edu/1996/012696/Star012696.html\\
&&&&&&&&http://www.washington.edu/research/pathbreakers/1994c.html\\
&&&&&&&&http://www.apo.nmsu.edu/Telescopes/eng.papers/performance/performance1988.html \\
Starfire 3.5m & AFRL & 3.5 & 27.0 & 1993 & 30.9 &   &  & http://www.de.afrl.af.mil/Factsheets/35meter.html \\
WIYN & WIYN Consortium & 3.5 & 14.0 & 1994 & 15.7 & 1.8 &  & http://www.noao.edu/wiyn/wiynis.html \\
AEOS & USAF (Maui) & 3.7 & 18.2 & 2000 & 18.2 & 1.5 & 75 & http://www.acec.org/programs/2000eeaawards.htm\\
&&&&&&&&http://ulua.mhpcc.af.mil/AMOS/1999\_AMOSTechnicalConference/Mayo\_paper/Mayo.html \\
VISTA & ESO & 4.1 & 51.5 & 2003 & 48.0 &   &  & http://www.schott.com/english/news/press.html?NID=1417 \\
SOAR & CTIO & 4.2 & 28.0 & 2001 & 27.4 & 1.8 &  & http://www.lna.br/soar/soar\_e.html \\
Magellan 1 & CfA & 6.5 & 65.0 & 2000 & 65.0 & 1.3 & 150 & http://www.unispace3.co.cl/inicial6\_e.html \\
Magellan 2 & CfA & 6.5 & 72.0 & 2001 & 70.4 & 1.3 & 150 & http://www.astro.lsa.umich.edu/$\sim$rab/specjust.pdf \\
Gemini & NSF & 8.1 & 88.0 & 1992 & 103.5 & 1.8 & 311 & http://www.the-scientist.com/yr1992/march/let3\_920316.html \\
VLT UTs & ESO & 8.2 & 266.8 & 2000 & 266.8 & 1.8 & 430 & http://www.phys.hawaii.edu/$\sim$jgl/post/WGLSF\_table\_29Mar01.htm\\
&&&&&&&&http://www.belspo.be/belspo/res/coord/res\_euro/eso/det\_en.stm \\
Subaru & NAOJ & 8.3 & 170.0 & 2000 & 170.0 & 1.8 & 500 & http://www.hawaii-county.com/databook\_97/section13.htm\\
&&&&&&&&http://www.phys.hawaii.edu/$\sim$jgl/post/WGLSF\_table\_29Mar01.htm\\
&&&&&&&&http://www.jinjapan.org/kidsweb/techno/subaru/history.html \\
Keck & CARA & 10.0 & 94.5 & 1985 & 139.1 & 1.8 & 273 & http://www.lbl.gov/Science-Articles/Archive/keck-telescope.html \\
GTC & Spain & 10.0 & 90.6 & 1997 & 95.3 & 1.7 &  & http://www.gtc.iac.es/home.html; 12750 MPtas (1997) \\
LBT & Univ. Arizona & 11.9 & 110.0 & 2000 & 110.0 & 1.1 & 530 & http://www.sdaa.org/SDAAAppli/arizona.htm\&e=747\\
&&&&&&&&http://medusa.as.arizona.edu/lbtwww/\\
&&&&&&&&http://mytwobeadsworth.com/MtGraham928.html \\

LMT & Univ. of British Columbia  & 6.0 & 1.0 & 2000 & 1.0 & 1.5 & n/a &  \\
HET & McDonald Obs. & 9.5 & 14.0 & 1997 & 14.7 & 1.8 & n/a &  \\
BTA & Special Astroph. Obs. & 6.0 &   & 1976 &   & 4.0 & 850 &  \\
UKIRT & UK & 3.8 & 5.0 & 1979 & 10.9 &  &  & http://www.hawaii-county.com/databook\_97/section13.htm \\
UH 88" & UH & 2.2 & 5.0 & 1970 & 20.8 &   &  & http://www.hawaii-county.com/databook\_97/section13.htm \\
MMT & Univ. Arizona & 6.5 & 49.4 & 2000 & 49.4 & 1.3 & 118 & \$11m for original 6 elements in 1977, \$20m conversion in 1995 \\

CFHT Upgrade  & CFHT Consortium & 8.0 & 108.0 & 1999 & 110.5 &   &  & http://www.casca.ca/lrp/vol2/wf8m/node5.html \\
Swedish ELT & Lund Obs. & 50.0 & 876.0 & 2003 & 816.8 &   &  & http://www.astro.lu.se/$\sim$torben/euro50/publications/swedish50m99.pdf; 750 million Euros \\
GSMT / CELT & NOAO / Caltech & 30.0 & 600.0 & 2000 & 600.0 &   &  & http://www.aura-nio.noao.edu/book/ch5/5\_7.html \\

\enddata
\end{deluxetable}


\begin{figure}\label{fig_cost_dia}
     \epsscale{1.0}
     \plotone{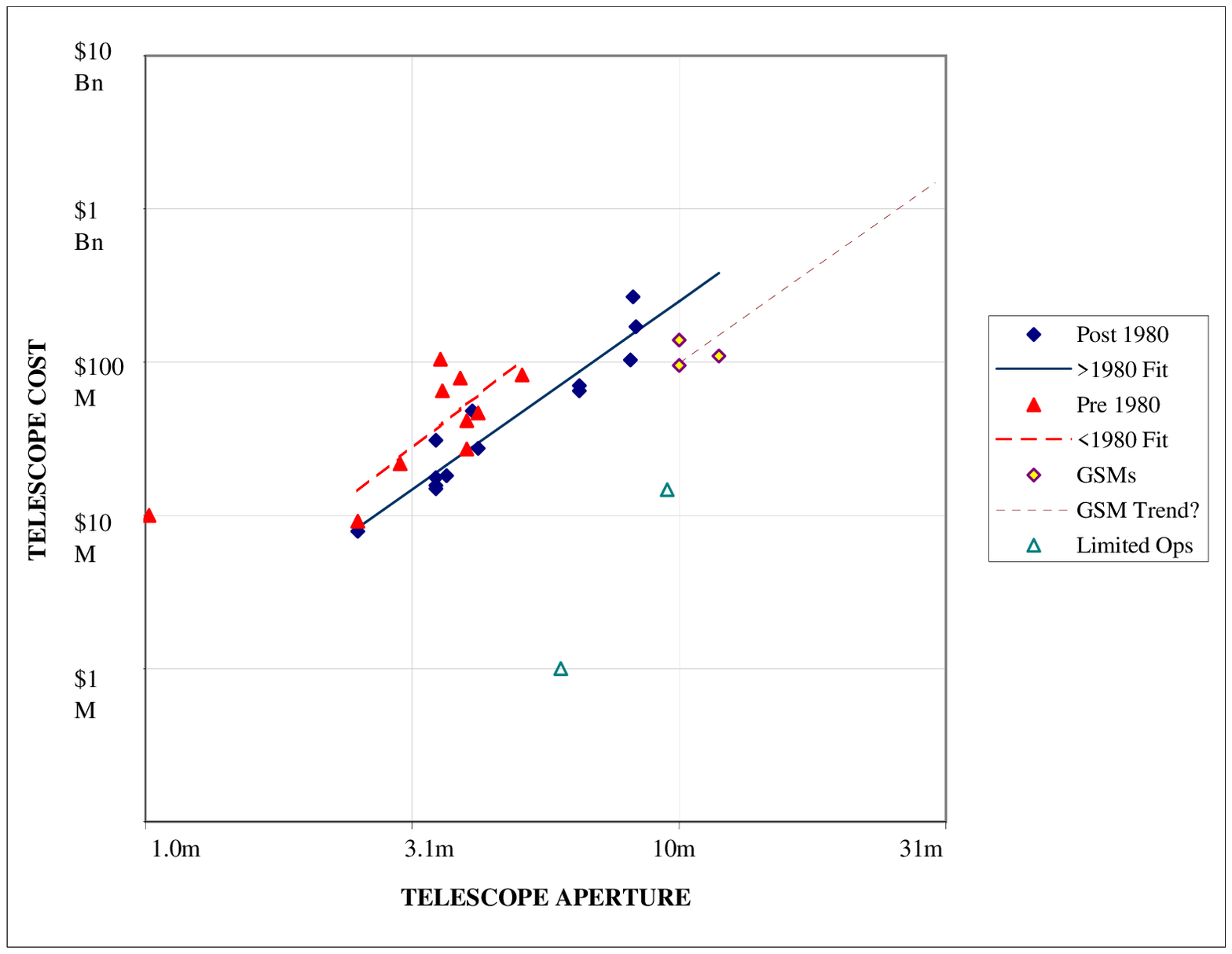}
     \caption{Cost versus aperture diameter for optical telescopes built before
     and after 1980.  For the pre-1980 fit, cost $\propto D^{2.77}$, and for the
     post-1980 fit (exclusive of the giant segmented mirrors), cost $\propto D^{2.45}$. The two
     limited operations telescopes plotted are the UBC 6-m liquid mecury telescope and
     the 9-m (effective) HET.}
\end{figure}

\subsection{Discussion on Ground-Based Apertures}

As seen in Figure 1, there appears to be a clear progression of cost with telescope size.
An examination of the aperture built prior to 1980 shows cost $\propto D^{2.77}$, which
as would be expected is consistent with \citet{mei78}.  For those monolithic apertures
built since 1980, the cost-aperture power law is slightly shallower, with
cost $\propto D^{2.46}$, but still significantly greater than merely scaling with telescope
area, $D^2$.  The GSM telescopes that have been built appear to drop below the post-1980
line, just as the post-1980 line drops below the pre-1980 line.

Our interpretation of this offset in the power law intercept is the cost reducing
impact of fundamentally new technologies.  At the $\sim$1980 turning point, the improvement was
a combination of telescope mounting and faster optical systems, reducing the overall
size of the telescopes.  For the advance associated with GSMs, the improvement is
the cost reduction associated with fabrication of segmented versus monolithic
primary mirrors.  There are unfortunately not enough data points to determine
if the GSMs will also follow a cost $\propto D^{2.46}$ power law; however, we naively
expect for the cost-aperture relationship to generally adhere to this slope.

As such, we
may easily predict general costs for future apertures built
using technology associated with the current family of GSMs.  We expect a 30-m
telescope to cost roughly \$1.4 billion, and a 100-m telescope is expected
to be roughly \$26 billion, {\it using current GSM technology}.  If, as can be reasonably
postulated, advances in telescope construction technology can be applied to
the next generation of large apertures, reductions of 2-3$\times$ can be expected with
each new family of technology, as seen in the progression from pre-1980 to post-1980 to
GSMs.  A \$600M, 30-m telescope can be reasonably argued to be only a single technology
generation away.  However, following this same reasoning, a \$2B, 100-m telescope is
probably a full three technology generations away from being realized.

\section{Space-Based Telescopes}

Unfortunately, there are only a few operational examples of space-based
telescopes.  The obvious candidate is the Hubble Space Telescope (HST).  Of NASA's other
three `Great Observatories', only the
Space Infrared Telescope Facility (SIRTF) has a mirror design that lends itself
to comparison within this context.  A full accounting of flight designs is appropriate
within the context of categorizing the approaches to space telescopes:

\begin{itemize}
\item Delivery to orbit - Both HST and SIRTF are examples of this category of space-based
mission.
\item Assembly in orbit - Given the payload shroud constraints of a $\sim$5m diameter on
even the largest of launch vehicles, a number of spacecraft that have flown or are
in the planning stages take advantage of a ground-based construction, with a space-based
assembly stage.  This can be as simple as an autonomous unfurling, or a more complicated
and drawn out assembly phase prior to operations.
It is worth noting that there are two obvious classes of telescope in this category - those
that benefit from {\it robotic} assembly, and (particularly within the context
of the space station) those that would be the product
of {\it human} assembly.  Of surprise to many, there are three clear examples of at least the
robotic assembly approach to
date: the 8-m VSOP and 12-m commercial MBSat radio antennas, both of which have flown, and
the 6.5-m near-infrared JWST, which has not flown but is
clearly committed to this approach and will be orbited within the next ten years.
\item Fabrication in orbit - This approach is, at present, somewhat more fanciful than the previous two,
potentially making using of some sort of {\it in situ} resource utilization (and as a result,
bypassing the limitations of launch vehicle lift restrictions).  Although the most promising in
terms of ultimate aperture size, we will only mention this approach here in passing, for the
sake of completeness, due to its gross technical immaturity.
\end{itemize}
A further complication worth considering is the prospect of {\it on-orbit servicing}, which
can be applied equally to all three categories above.

Given the small number of examples to date for space-based telescopes, no general inference can
be drawn from the relationship between telescope cost and aperture size for these apertures.
Indeed, the similar relative cost between Hubble and JWST - on the order of \$1 to \$2 billion
dollars for each - would indicate within the simple confines of the
rough analysis presented herein for ground-based apertures
that telescope size is independent of cost.  Instead,
our assessment is that the predominant phenomenon at play is rapid technological development
as it impacts aperture size, rather than simple scaling of a single family of technology.

\section{General Discussion}


{\it Ground-based Telescopes.} There are two key factors that affect the aperture-cost scaling
law for ground-based telescopes:
\begin{itemize}
\item Environment - Environment manifests itself in two significant
ways for ground-based telescopes.
Inclement weather is the first of these two ways - the telescope must be protected from
precipitation and other hazards associated with being open to the air.  For all major optical
telescopes to date,
this is accomplished by construction of a telescope enclosure, typically
a dome.  The second weather factor is wind - acceptably low wind velocities do not preclude
operation under transparent conditions, but wind shake can significantly degrade telescope
performance.  For most optical telescopes, an enclosure can also mitigate the effect
of wind upon the telescope structure, typically a co-rotating dome.

For optical
telescopes, as the aperture grows, the dome grows as $\sim (\textrm{f/ratio} \times D)^3$.  Reduction of telescope
f/ratios over time have improved the situation over the past twenty years, but
this factor ultimately can be no smaller than $\sim D^3$.
A common mistake at many observatories is the assumption that the dome is a simple
element of the overall observatory, not worth a great deal of thought or investment; the
result is often years of expensive maintenance headaches and/or operational limitations.

Recent illustrations accompanying proposals for a 30m-class telescope compare the aperture size
to that of a baseball diamond; this is a particularly illustrative example, noting that
recent retractable roof baseball stadiums have been built and are worthwhile
enclosures to consider when trying to approximate price.  While larger than the
enclosure for a 30m telescope (in that they have to enclose an outfield and
grandstands), they are also significantly simpler in that they only retract, and do not
have to rotate.  A recent example of this sort of venue in Seattle was built for \$600M
(telescope not included).

\item Gravity - Observational pointing access to the sky is typically achieved through orienting
the telescope structure in two axes, frequently elevation \& azimuth or right ascension \&
declination.
(The Hobey-Eberly and Liquid Mercury Telescope are notable exceptions to
this observation, and have traded significant operational flexibility for
economic advantage, as seen in Figure 1.)  Changing a
telescope's elevation or declination alters the angle
between the telescope's pointing vector and the local gravity vector.  Since the telescope
must maintain its alignments throughout all pointings, the structure must be tolerant
of this variable angle.  As such, the telescope structure often grows as a hemisphere
behind the aperture it supports - the growth, and cost, of this structure will scale
as $\sim D^3$.  Clever design of this structure can reduce the power law to something
closer to the square of the aperture diameter, but consistency of the
$D^{2.7}$ aperture-cost scaling law indicates there are perhaps limits to cleverness
dictated by modern construction materials and techniques.
\end{itemize}

Both of these factors affect the relevant power law for ground-based telescopes.
Elimination of the telescope dome for the largest of the new telescopes is certainly an option
(and actively under consideration for some of the larger apertures proposed),
although it will clearly multiply the deleterious effect of wind shake on the telescope
backing structure and push the cost of the backing structure back towards $\sim D^3$.
These two ever present ground-based factors will push the aperture-cost scaling law away
from $\sim D^2$ and towards $\sim D^3$.

{\it Space-based Telescopes.}
As with ground-based apertures, there are two key factors that affect the aperture-cost scaling
law for space-based telescopes:

\begin{itemize}
\item Structural stability - As with ground-based telescopes, a space-based telescope's backing
structure will be responsible for maintaining the unique shape of the primary mirror, regardless
of pointing.  However, given the absence of a significant gravitational field, the
structure may be designed primarily for aperture alignment rather than support against an
external field.  There will be no changing external force to cope with as the aperture points
to different portions of the sky.  As such, it is our expectation that the
structure will be primarily 2-dimensional assembly and that the
aperture-cost law associated with maintaining optical figure will scale as $\sim D^2$.

\item Environment - For structures of significant size in space, an important consideration
that begins to impact operational considerations is the space `weather', primarily
due to the sun.  Particulate solar wind, radiation pressure, and heating effects
of the solar environment will all have to be accounted for.
It is likely that large telescopes in space will need a shield between the primary aperture
and the sun.  This shield,
while notionally as large or even larger than the aperture itself, will also manifest
itself as a primarily 2-dimensional structure.  Also, given the substantially relaxed
requirements for such a shield to maintain a given shape, it can be a fairly
gossamer structure.  Such a shield provides the additional benefit of cooling
of the telescope; this type of structure is already a part of the baseline JWST design
and is not considered to be a significant cost driver.
The cost of this structure should also scale as $\sim D^2$.
\end{itemize}

It is also worth noting that certain expensive design drivers for ground-based telescopes are
not necessarily present in punitive space-based designs.  For example, since a dome is no longer
enclosing the telescope structure, a driver for relatively fast focal ratios (and difficult
to fabricate parabolas) is removed.

Overall, our expectation is that ground-based telescope costs will continue
to scale as $\sim D^{2.5}$.  Improvements in technology will provide one-time
shifts in the zero-point of the aperture-cost relationship, with no impact
upon slope.  In contrast to the ground-based case, we expect space-based
apertures to have a much slower aperture-cost relationship, growing as
slowly as $\sim D^{2.0}$.  The difference in slopes has a striking consequence:
{\bf At some given aperture size, it will be just as expensive to deliver an operational
space-based or ground-based telescope}.  This equality is {\it independent} of the
obvious advantages a space-based aperture has over its ground-based counterpart.
Isolating the cross-over point of the two power laws will be of particular interest,
in that it points to the size domain that will be exclusively inhabited
by space-based apertures.

At the present, using these putative values for the power law slopes, and starting
from the points established by the current generation of GSMs for the ground-based case,
and JWST for the space-based case, the cross-over point appears to appear at the
300m filled aperture size, at a cost of \$100 billion dollars.  This is a completely
unrealistic sum for any telescope.  However, if we advance from this starting point
and move forward two technology generations for both space-based and ground-based telescopes -
with the attendant shift in power law intercepts - our cross-over location shifts to
a 120m filled aperture at a cost of \$10 billion dollars.  This is still quite a speculative
sum, but getting to be significantly more realistic.  If there is a more rapid advance
in space technology than in ground-based telescope technology (which these authors
do not think untenable given the relative levels of investment),
and two generations of space-based observatory technology evolve
for every one of on the ground, our cross-over shifts to 70-m, \$3 billion dollars.  Given
these sorts of possible scenarios, it is our expectation that the largest aperture built upon
the ground will be in the region of 100-m.

Above and beyond the initial cost of an observational facility,
there are two additional aspects of telescope finances that are not being examined in great detail
in this simple analysis, but they bear mentioning here:
\begin{itemize}
\item Instrumentation - A substantial portion of the cost of any operational observatory is
its instrumentation.  For ground-based apertures, this can be an evolving suite of instruments
with various specialized specifications and design goals.  For these facilities, and for those
space-based observatories with on-orbit servicing, ongoing instrumentation upgrades represent an
ongoing cost of the facility.

\item Operational Costs - For ground-based observatories, this number can run annually from 5\% to 30\%
of the overall initial construction cost.  There are two aspects of this cost that
can be specifically identified here: first, that of ongoing maintenance, and second, that of
the actual observing done with the facility.
\end{itemize}
For those space-based observatories that do not benefit from on-orbit servicing, some of these
costs simply do not appear - new instrumentation does not need to be developed, nor does daily
maintenance need to be physically performed upon the spacecraft(s).  However, this potentially translates
into limitations in terms of instrument capability and mission lifetime, particularly in relation
to ground-based facilities, so the actual benefit or penalty of these considerations is
not entirely clear.

\section{Conclusions}

We have shown the telescope cost growth scaling law of $\sim D^{2.77}$ that was first
noted in \citet{mei78} for ground-based telescopes is slightly shallower
for the apertures that have been built since 1980, at $\sim D^{2.46}$,
but remains generally true.  We have also
presented arguments in support of a similar, but notably shallower, scaling law for space-based
telescopes, closer to $\sim D^{2.0}$.  An important implication of these two power laws is their
intersection - this point defines a telescope that will be equally expensive to build on the
ground or in space.  This point is independent of the advantages to be gained in siting
the aperture in space versus on the ground.

This is particularly interesting as the astronomical community contemplates construction
of ground-based apertures that are up to 100m in size.  Given the limits of public and private
support for construction of new telescopes, it would be prudent for the community to carefully
consider the directions they take their more ambitious technology development efforts.  These
investments should be directed with consideration regarding if
those efforts eventually will dead end, as in the ground-based case, or have
substantial growth options.
Current thinking in terms of
`overwhelmingly' large apertures will of course eventually give way to thoughts about
even larger instruments capable of achieving science goals with fundamental implications
for astrophysical discovery.
These goals include continental mapping of
nearby exosolar terrestrial planets, a complete sub-pc catalog of star formation within the
local group, and surveys of the early universe at $z>10$.

\acknowledgments

We acknowledge fruitful discussions with a large number of individuals in the field,
most of whom expressed varying degrees of healthy skepticism at our premise.
Portions of this work were
performed at the California Institute
of Technology under contract with the National Aeronautics and
Space Administration.


\begin{thebibliography}{}
\bibitem[Allen(1973)]{all73}Allen, C.W., 1973, Astrophysical Quantites, London: University of London, Athlone Press, 3rd ed
\bibitem[Meinel(1978)]{mei78}Meinel, A.B., 1978, in Optical Telescopes of the Future (eds Pacini, F., Richter, W., Wilson, R. N.) 13-26 (ESO Conf. Proc.)
\bibitem[Meinel(1979a)]{mei79a}Meinel, A.B., 1979, Proc. SPIE 172, 2
\bibitem[Meinel(1979b)]{mei79b}Meinel, A.B., 1979, Optical Engineering, vol 18, no 6, 645
\end{thebibliography}
\end{document}